\documentclass[12pt]{article}
\usepackage[margin=1in]{geometry}
\usepackage[utf8x]{inputenc} 
\usepackage[english]{babel}
\usepackage{amsmath,amsthm}
\usepackage{mathtools}
\usepackage[ruled]{algorithm2e}
\usepackage[dvipsnames]{xcolor}
\usepackage{color}

\theoremstyle{definition}
\newtheorem{definition}{Definition}

\theoremstyle{plain}
\newtheorem{theorem}{Theorem}
\newtheorem{lemma}{Lemma}
\newtheorem{corollary}{Corollary}

\begin{document}
\title{Communication Lower-Bounds for Distributed-Memory Computations for Mass Spectrometry based Omics Data}
\author{Fahad Saeed\footnote{Corresponding Author}, Muhammad Haseeb, SS Iyengar \\School of Computing and Information Sciences (SCIS)\\ Florida International University (FIU)\\ Miami FL USA 33199\\ Email: fsaeed@fiu.edu}

\maketitle

\begin{abstract}
Mass spectrometry (MS) based omics data analysis require significant time and resources. To date, few parallel algorithms have been proposed for deducing peptides from mass spectrometry-based data. However, these parallel algorithms were designed, and developed when the amount of data that needed to be processed was smaller in scale. In this paper, we prove that the communication bound that is reached by the \emph{existing} parallel algorithms is $\Omega(mn+2r\frac{q}{p})$, where $m$ and $n$ are the dimensions of the theoretical database matrix, $q$ and $r$ are dimensions of spectra, and $p$ is the number of processors. We further prove that communication-optimal strategy with fast-memory $\sqrt{M} = mn + \frac{2qr}{p}$ can achieve $\Omega({\frac{2mnq}{p}})$ but is not achieved by any existing parallel proteomics algorithms till date. To validate our claim, we performed a meta-analysis of published parallel algorithms, and their performance results. We show that sub-optimal speedups with increasing number of processors is a direct consequence of not achieving the communication lower-bounds. We further validate our claim by performing experiments which demonstrate the communication bounds that are proved in this paper. Consequently, we assert that next-generation of \emph{provable}, and demonstrated superior parallel algorithms are urgently needed for MS based large systems-biology studies especially for meta-proteomics, proteogenomic, microbiome, and proteomics for non-model organisms. Our hope is that this paper will excite the parallel computing community to further investigate parallel algorithms for highly influential MS based omics problems. 
\end{abstract}

\section{Introduction}
Almost all numerical algorithms when developed, considered \emph{arithmetic operations} as the sole metric for efficiency \cite{ballard2014communication}. Over time, especially in the last decade, the technological trend of the Moore's law has kept making the arithmetic operations faster. Therefore, bottleneck for many algorithms have shifted from computational arithmetic operations efficiency to \emph{communication} i.e. communication costs of moving the data between different memory-distributed processors connected via a network. Communication of data elements is essential because operand requires them to be in the same memory at the same time. Same applies to serial machines where the data has to be moved to the smallest, and fastest memory in the hierarchy (i.e. cache). Numerous studies have shown \cite{ballard2011minimizing,national2005getting,ballard2014communication} this trend of excessive cost of moving data exceeds the costs of doing the arithmetic operations. With the introduction and ubiquitous multicores, manycore, and GPU based architectures; this gap is, and will continue to grow exponentially over time \cite{demmel2013communication,solomonik2011improving}. The current trend for increasing the efficiency for most numerical algorithms is to reduce the gap between moving the data, and computing on that data. This trend is observed both for serial as well as parallel algorithms \cite{ballard2014communication}.

Over the years, significant efforts have been invested for the design, and development of efficient methods for Mass Spectrometry based omics data analysis. These \emph{numerical algorithms} include highly successful search engines including but not limited to Sequest \cite{eng2008fast,diament2011faster,eng1994approach}, Tide \cite{mcilwain2014crux}, Mascot, XTandem, and more recently MSFragger \cite{kong2017msfragger}. Recent trends in systems biology Mass Spectrometry (MS) based experiments generate increasingly large, and complex (multiple species, non-model species, microbiome) data sets potentially leading to high-impact proteomics, meta-proteomics and microbiomes studies directly related to human disease and health. This in turn point towards need for larger, better, and faster computational tools \cite{doi:10.1021/acs.jproteome.8b00761,awan2017out,haseeb2019lbe}. The numerical algorithms developed for Mass Spectrometry (MS) based peptide deduction are designed and implemented by assuming number of \emph{arithmetic operations} as the sole metric for efficiency.

Development of computational methods for MS data is an active area of research \cite{kong2017msfragger} but the focus of this research, till date, has been towards improving the efficiency of arithmetic operations. Database-search workflows are the most used data processing pipelines which require matching a high-dimensional noisy MS data (called spectra) to a database of protein sequences. These MS data sets are then processed using databases which may be several times larger than the original proteome (or multiple proteomes in case of meta-proteomics studies \cite{yates2019proteomics}) depending on the search parameters. The data volume can easily reach tera-byte level depending on the experiment, and search parameters for these workflows. Non-model organism proteomics is considered the next frontier \cite{yates2019proteomics,heck2020proteomics} to accelerate insight into chronic disease in humans, and requires even larger search-spaces which would lead to intractable run-times. 

Increasing size of the spectra, and theoretical database search-space has led to the development of high-performance computing (HPC) strategies \cite{kulkarni2009scalable, li2019mctandem, sun2014improved, duncan2005parallel,bjornson2008x, li2019sw, li2019mctandem} to speed up these search engines. Similar to serial numerical algorithms, the objective of these HPC methods have been to speed up the arithmetic scoring part of the search engines with little to no efforts to minimize the communication costs. 

Exclusion of communication costs as a metric, of otherwise highly successful methods \cite{eng2008fast,diament2011faster,eng1994approach,mcilwain2014crux, kong2017msfragger}, is becoming a severe bottleneck for processing of MS data (due to excessive processing times), and now hinders scientific advancements for mass spectrometry, and (meta) proteomics/microbiome research. This observation might be common perception for systems biologist working with MS data analysis for meta-proteomics, proteogenomic, or microbiome studies but it is still anecdotal. In other words, it is hard to quantify the good or poor scalability of these workflows. From our anecdotal observations, it was apparent that existing parallel algorithms result in abysmal speedups with increasing number of processors or data sets. To prove that these observations are not an artifact of a specific library or compute-architecture we wanted to investigate the lower-bounds that are acquired by existing HPC algorithms. 

Therefore, we set out to ask these questions: (1) Are there lower-bounds on the parallel algorithms that can be acquired? (2) Do existing HPC algorithms attain these lower-bounds? (3) If not, are there new parallel algorithms that will allow us to do that? The bounds will be similar for serial algorithms subject to architecture-specific communication costs. To date, we are not aware of compute- or communication bounds proved for any MS based omics serial or parallel algorithms.

In this paper, we theoretically prove that the efficiency of these algorithms (both serial and parallel) are bottle-necked by the communication costs, and are prohibitively excessive; no matter what kind of indexing is employed. We further prove the theoretical lower-bounds that \emph{are} possible but are not achieved by any existing parallel algorithm. Lastly, we demonstrate that these lower-bounds were consistent with the empirical observations (and published results), and experimental resultsAs expected, attaining these lower-bounds would require a significant redesign of these parallel (and serial) algorithms; and not just OPENMP loop transformations. These redesigns may include different numerical properties, transformation of MS data into readable/write-able compressed formats, more effective ways of decomposing the data on parallel architecture that incur minimal communications, and different data-structures that need to be investigated.

Rest of the paper is organized as follows. In section \ref{communication-model}, we formulate the communication models that would be used for analysis of parallel algorithms. In the next section \ref{proteomics-workflow}, we introduce the reader to proteomics workflows, and a generalized parallel strategy that is used by all HPC methods. In section \ref{lower-bounds}, we provide theoretical proves of the communications bounds, the computation bounds, and the overall runtime bounds of the existing, as well as communication-optimal parallel algorithms. In section \ref{meta-analysis}, we provide the meta-analysis of all existing HPC methods, and analyze the published results with our new communication/computation bounds. In section \ref{experiments}, we elaborate on the experimental set up and details about the results that are obtained. Section \ref{discussions}, and section \ref{conclusions} are reserved for discussion and conclusions. 

\section{Communication Model} \label{communication-model}
For design of parallel algorithms, it is essential that they are not only load-balanced but also minimize the communication costs between processors associated with data decomposition. Most of the algorithms, especially ones dealing with big data sets, have inter-processor communications costs that are much larger than the computation costs. Hardware trends that are growing towards more many-core, and multi-core architectures also predict that most of the problems will become communication-bound even for serial algorithms \cite{ballard2012communication}. 
For our MS based proteomics parallel algorithms, we will model the cost of communications as follows: There are two costs that are normally associated with communication. When the system has to send $n$ words from one processor to the other over the network via which the processors are communicating; the words are first packed into contiguous block of memory and is known as a \emph{message}. This message is then sent to the destination processor by following the parallel algorithmic constructs that have been implemented. There is a fixed overhead time that is required to assemble, pack, and transmit the data (called latency cost denoted by $\alpha$). There is also time needed to transmit $n$ words and this time is proportional to $n$ called the \emph{bandwidth cost} denoted by $\beta n$. Then to send one message of $n$ words is denoted by $\alpha + \beta n$, and the time to send $S$ messages containing a total of $W$ words can be denoted by $\alpha S + \beta W$. Also let $\gamma$ denote the time it takes to perform one arithmetic computation, and $F$ denotes that total number of computations. Summation of all of these terms is equivalent to $\alpha S + \beta W + \gamma F$, and the recent technological trends dictate that $\alpha >> \beta >> \gamma$. Therefore, it is of utmost importance to have parallel algorithms that can minimize \emph{both} the bandwidth, and the latency. Such communication models are used for minimizing communication in numerical linear-algebraic computations, and more details can be found at \cite{ballard2011minimizing}.

\subsection{Sequential Computer}
For a serial architecture that has levels of memory-hierarchy, the model $\alpha S + \beta W + \gamma F$, would suffice for 2 levels of hierarchy. If there are more levels are hierarchy to be considered then there is a communication cost associated with each level and when the data is moved to/from that level.

\subsection{Parallel Computer}
Similar to a sequential computer, 
$\alpha S + \beta W + \gamma F$ would be sufficient to provide the communication costs associated with \emph{one} node of the parallel computing architecture. A \emph{lower-bound} on one processor is enough to get a lower-bound on the whole algorithm with the assumption that all processors are homogeneous and are completing the same tasks. A \emph{upper-bound} (time required by the entire algorithm) will need a summation of the terms in an order of dependencies considering the critical path, which maximize the summation of these costs. If the parallel architecture can overlap communication and computations; then the expression can be replaced with 
$max (\alpha S + \beta W, \gamma F) $ or 
$max (\alpha S, \beta W, \gamma F)$ which can lower the cost by 2 or 3 but does not effect the asymptomatic relations. Different indexes can be used for formulating the model for a heterogeneous architecture. However, for this paper we will assume a homogeneous architecture. 

Finally, an algorithm will be called a \emph{communication-optimal} algorithm if it can asymptotically attain the communication lower-bounds for a given parallel architectures. Such an algorithm is also colloquially known as \emph{communication-avoiding}. 

\section{MS Database Proteomics, Proteogenomic, and Meta-Proteomics Search} \label{proteomics-workflow}
\begin{figure*}[h] 
\centering
\includegraphics[scale=0.75]{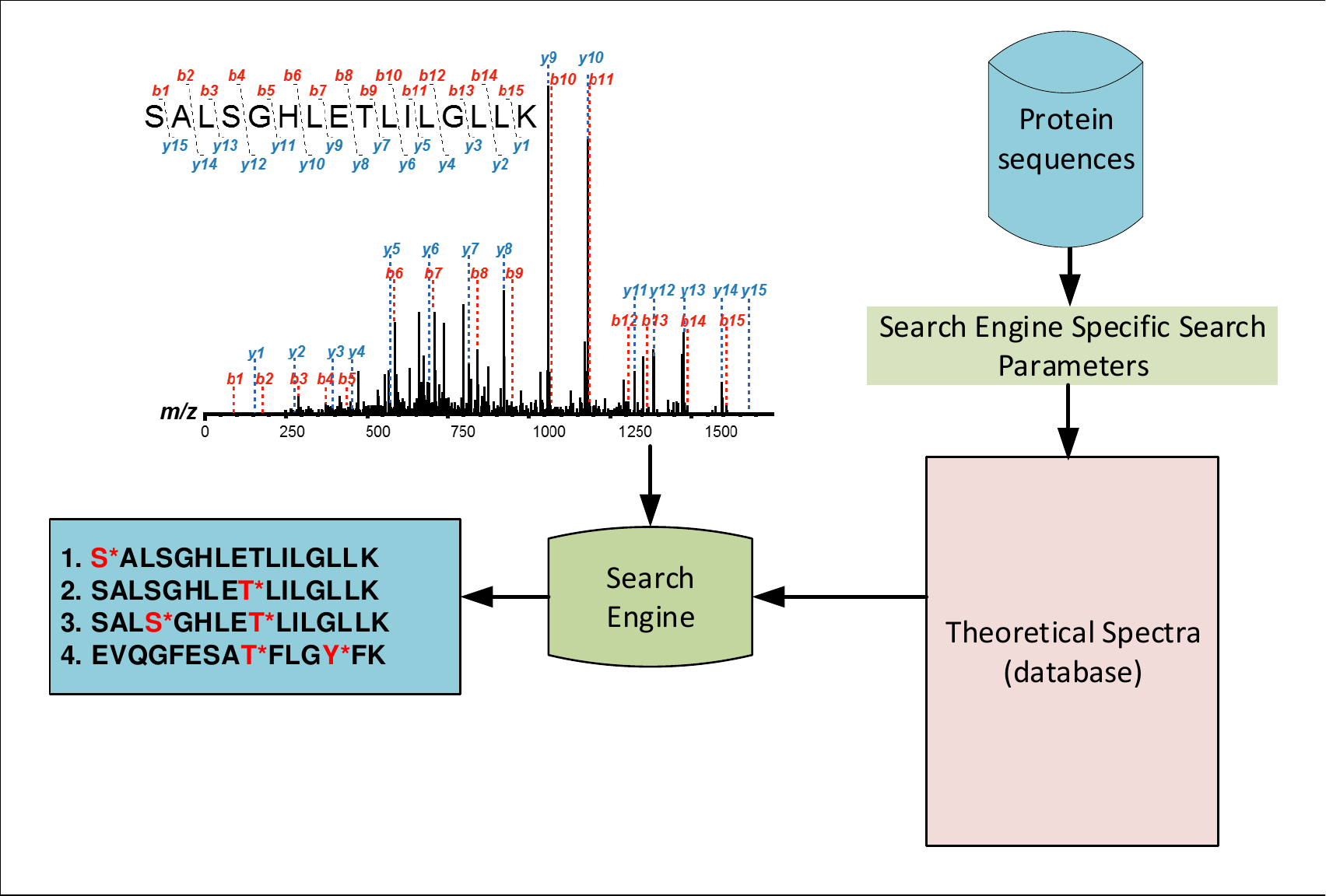}
\caption{A high-level overview of the MS based proteomics data analysis that leads to spectra-to-peptide deductions.}
\label{workflow}
\end{figure*}

We will start by defining the \emph{database-search} strategy that is used for Mass spectrometry data. For the purposes of this framework we will assume the most simplest strategy independently on how the data was acquired and what are the systems biology objectives. This will ensure that our results are generalised for most of MS data processing using databases. The most commonly employed method for peptide identification is the database search where the experimental tandem MS/MS spectra are compared to the theoretically predicted spectral libraries/databases \cite{kong2017msfragger}. The theoretical spectral libraries are generated by first \emph{in-silico} digesting a proteome sequence database into peptide sequences and then predicting MS/MS spectra for each peptide sequence and its possible (modified) variants. The advantage of this technique is that Post-Translational Modifications (PTM) and fragmentation types can be easily incorporated in the theoretical spectra. The experimental spectra is then compared with the theoretical spectra created during the database creation process just described. This scoring is called peptide-to-spectrum match (PSM) computations. An overview schematic of the mass spectrometry based peptide deduction is shown in Fig. \ref{workflow}.

\subsection{Generalized Parallel Computing Strategy}\label{general-hpc}
Existing parallel algorithms for proteomics, like numerical algorithms in other domains, have been designed for problems that are compute-bound. In general, all HPC algorithms that have been proposed in this domain operate by taking the database and distributing it over the processors. Once the database is communicated, $N/p$ of the spectra is assigned on each processing unit where $N$ in the total number of spectra and $p$ is the number of processing elements. Thereafter, a serial algorithm (such as XTandem) is executed on each of the node in parallel. Once this is completed the results are transmitted back to the master node. It is easy to generalize these HPC methods and are listed in Algorithm 1. Note that in these methods few assumptions are made that may not be true for today's calculations i.e. each spectra takes equal amount of computations, the communication is minimal and the overall workflow is compute-bound. Therefore, no significant effort is invested in getting a load-balanced system, or minimizing the communication costs. We show in this paper that both of these factors are now a major bottleneck for these parallel algorithms. 

\begin{algorithm}[H]
\SetAlgoLined
\KwResult{Each Spectra is assigned to a peptide}
 \While{Spectra need peptide deduction}{
 \begin{enumerate}
  \setlength{\itemsep}{0pt}
  \setlength{\parskip}{0pt}
  \setlength{\parsep}{0pt}
  \item Take a species specific protein database; and expand it to a theoretical database D using search parameters\;
  \item Database D is copied whole on each of the P processors\;
 \item The spectra set S that needs to be processed are divided in S/P parts\;
 \item S/P spectra are processed on each of the processor in parallel\;
 \item The results are accumulated using MPI-gather or similar operation\;
 \end{enumerate}
 }
\caption{General HPC strategy that is used by Parallel Methods for MS based Proteomics data}
\end{algorithm}

\section{Communication Lower Bounds} \label{lower-bounds}
We will formulate the problem in terms of matrix operations, and prove the computation- and communication bounds for the existing strategies.

\begin{definition}
Database is the result of the theoretical spectra that are generated using the search parameters. Let this database be presented as a $m \times n$ matrix D where $m$ presented the number of theoretical spectra entries, and $n$ presents the average length of the entries. The entries of matrix D can be access using $i$, and $j$ indexes where $(0 \leq i < n)$, and $(0 \leq j < m)$. Then rows of D can be access using $D(0,j), D(1,j)$ and so on. \end{definition}

\begin{figure*}[h] 
\centering
\includegraphics[scale=0.5]{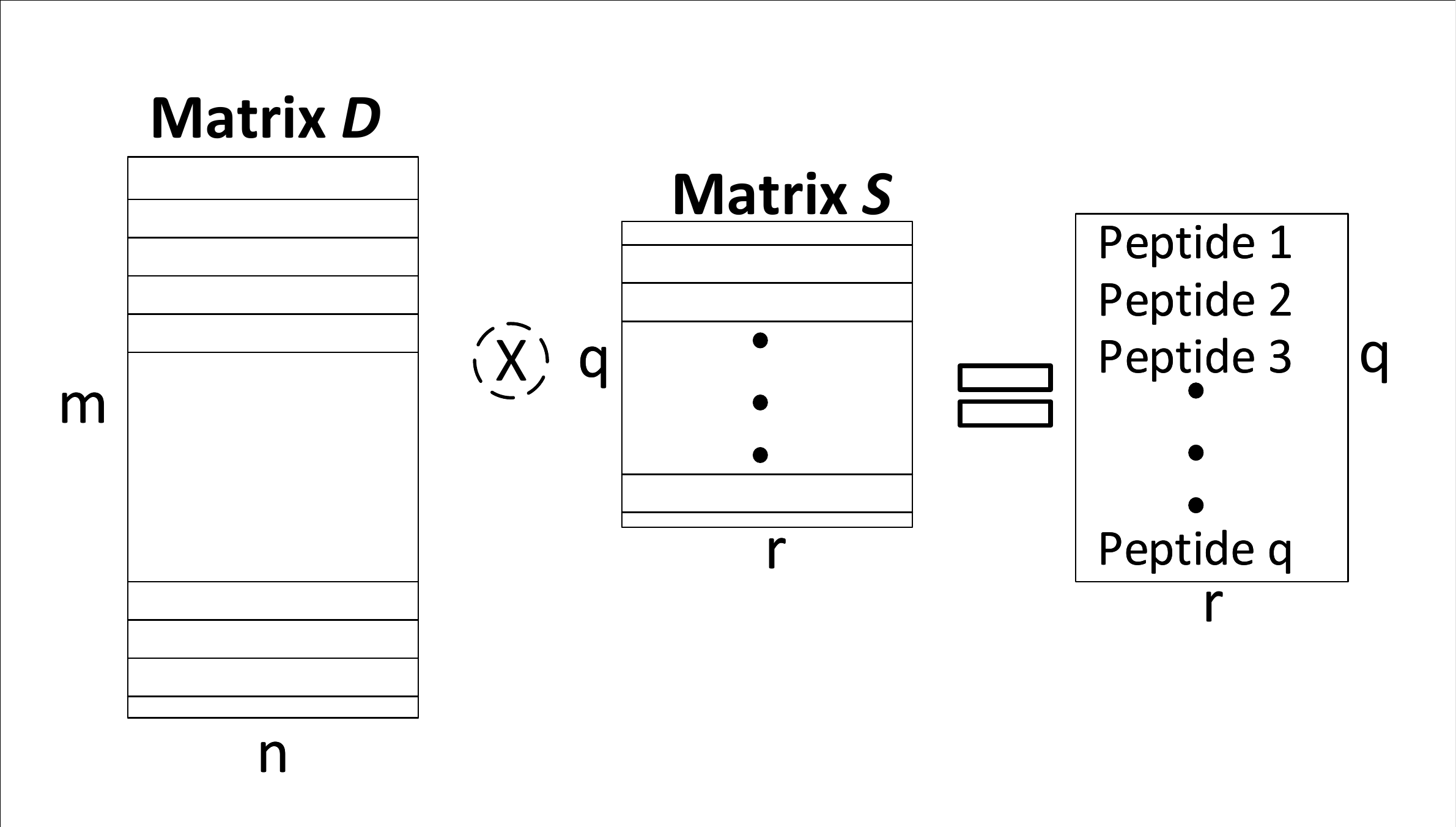}
\caption{A schematic of the matrix D (that represents the theoretical spectra), matrix S (that represents the experimental spectra), and matrix that holds the peptide that are deduced.}
\label{sch-fig}
\end{figure*}

\begin{definition}
Let the \emph{set} of spectra $s_0,s_1, \cdots, s_{(q-1)}$ that needs to be processed can be presented by a matrix S $q \times r$ where $q$ represents the number of spectra, and $r$ represents the average length of the spectra. The entries of matrix S can be access using $i$, and $j$ indexes where $(0 \leq i < r)$, and $(0 \leq j < q)$. Then rows of S can be access using $S(0,j), S(1,j)$ and so on. 
\end{definition}

A rough schematic of the matrix D, Matrix S, and the deduced peptides is shown in Fig. \ref{sch-fig}.

\begin{definition}
The parallel architecture is a memory-distributed processors with $M$ fast-memory associated with a single-core processor. All processor are assumed to be connected to each other. 
\end{definition}

\begin{lemma}
Three communication rounds take place for the existing parallel algorithms (similar to Algorithm 1) for MS based proteomics database search methods.
\begin{proof}
One communication round is the distribution of the database on each of the processors. Second communication round is the distribution of $q/p$ spectra on each processor. Third communication round takes place when the processing is done and the results of $q/p$ spectra are accumulated on a single machine.
\end{proof}
\end{lemma}

\begin{theorem}
The total words that are communicated using three round listed above are equal to $\Omega(mn)$ for existing HPC strategies.
\begin{proof}
The total words communicated on each processor is equal to $|D|+\frac{|S|}{p}+\frac{|S|}{p}$. Here it is easy to see that $|D| = (m \times n)$. Further $\frac{|S|}{p}$ is going to be equal to the words that are communicated from the spectra set i.e. $\frac{q\times r}{p}$. The final communication round is when the peptides are deduced for each spectra and accumulated on a processor. The words that will be communicated is equal to $\frac{q\times r}{p}$. $r$ is assumed to be the case where the spectra peaks are equal to the peptide length. Then the total number of words that are communicated is equal to $(m \times n)+\frac{q \times r}{p}+\frac{q \times r}{p}$ $= (m \times n)+2\frac{q \times r}{p}$. Therefore, the words communicated is $\Omega(mn+2r\frac{q}{p})$. 
\end{proof}
\end{theorem}

\begin{theorem} \label{comp}
The computational costs of dot-product like scoring that is performed for spectra-to-peptide match for each processor is equal to $F = \frac{qm(2n-1)}{p}$.

\begin{proof}
Each scalar dot product (called score) will work on one array from the database D and one array from the spectra S. On processor P0 which contains the whole matrix D, and subset of matrix from S; a score is calculated for $D(0,i)$ $0 \leq i \leq n$, and $S(0,j)$ $0 \leq j \leq r$. This will require $n$ multiplications, and $(n-1)$ additions. Since this has to be done for all entries of the database D; it will require $m \times (2n-1)$ computation for a single spectra. It is obvious that the number of spectra on each processor is $q/p$. This implies that the words that need to be processed on each processor is $\frac{qm(2n-1)}{p}$. 
\end{proof}
\end{theorem}

\begin{theorem}
The lower-bound of Bandwidth communication for database spectra to peptide match is W= $\Omega(\frac{m}{p})$ for any configuration of database or spectra in which dot-product scores are performed for matching.
\begin{proof}
The lower-bound of communication possible is equal to $\Omega(\# of Flops / \sqrt{M})$. The computations required for dot-product like routines is $O(\frac{qm (2n)}{p})$ as proved in our earlier theorem. The size of the fast memory is assumed to contain both the database, the spectra that needs to be searched and the result of the scoring. Therefore, $\sqrt{M} \geq mn + \frac{2qr}{p}$. Then the equation 
$\Omega{(\frac{2qm}{p \times (mn+\frac{2qr}{p})}})$. Our earlier assumption that $n \approx r$ and $q$ can be approaching $m$ is applicable here without losing generality. This gives us $\frac{m^2n}{pmn+2qn}$ which is equivalent to $\frac{m^2}{pm+2q}$. For $M$ which can contain the database, the spectra, and the results; As before for $q \approx m$ proves that lower-bound of communication which can be reached is equal to $\Omega{(\frac{m}{p})}$.  
\end{proof}
\end{theorem}


\begin{theorem}
We prove that the lower-bound on the Latency cost $L = \Omega(\frac{2}{mpn^2})$

\begin{proof}
A lower bandwidth bound on the bandwidth cost $W$ gives us a lower bound on the latency cost $L$. Assume that the largest message by a given architecture is $m_{max}$, then it is clear that $L \geq W/m_{max}$ since no message can be larger than the memory. Therefore we get $L = \Omega{(\frac{\# of flops}{M^{3/2}})}$. Assuming that $q \approx m$ the $\# of flops = \frac{m^2 (2n-1)}{p}$ then 
$L = \Omega{(\frac{m^2 (2n)}{p M^{3/2}})}$. Since we know that $\sqrt{M} = mn + \frac{2qr}{p}$; substituting will give us $L = \frac{m^2 (2n)}{p \times {(mn+\frac{2qr}{p}})^3}$. Since for large data sets $q \approx m$, and $n \approx r$; the expression can be approximated as $\frac{2}{mpn^2 \times (1+\frac{4}{p}+\frac{12}{p^2}+\frac{8}{p^3})}$. Therefore, $L \approx \Omega{(\frac{2}{mpn^2})}$. 
\end{proof}
\end{theorem}

\begin{theorem} \label{existing}
The overall runtime lower-bound of existing HPC methods is $\Omega{(mn)}$ irrespective of how many processors are used for computations.
\begin{proof}
The overall run time bound can be calculated for existing HPC methods can by summation of $L$, and $F$, and the communication that is specific to existing algorithms. The summation of these $\frac{qm(2n-1)}{p}$+ $(\frac{2}{mpn^2})$+ $(mn)$ gives us a lower bound on the overall run time which is bounded by $\Omega{(mn)}$.
\end{proof}
\end{theorem}

\begin{corollary}
Mass filtering (or other filtering specific to MS data) for candidate generation does not change the communication bounds of $\Omega{(mn)}$ of the current parallel algorithms.

\begin{proof}
Our communication bounds are proved by assuming that no mass filtering is taking place for computations. This is to ensure that the results are as genreralizable to parallel algorithms as possible; without considering specific algorithms. However, below we show that even with mass-filtering, communication bounds remain unchanged:

\emph{Case 1: The mass-filtering takes place on the master-node and the database, and truncated databases are communicated}
In the above case, the worst-case communication bounds is still going to be $\Omega{(mn)}$ since all (or a constant factor) of the database could be communicated at certain nodes. With the assumption that the parts that are transmitted are are fraction of the number of processors i.e. q/p; it is easy to see that the $\Omega((q/p)*mn)$ computations are needed for decisions at the master-node. Therefore, the communication bound remains unchanged. 

\emph{Case 2: The mass-filtering takes place on each node in parallel.}
If the mass filtering takes place on each node in parallel; then it needs to communicate $\Omega{(mn)}$ database to each node, and the communication bounds calculated in this paper remain unchanged. 
\end{proof}
\end{corollary}

\begin{corollary}
Fragment-Ion Index (based on MSFragger) scoring does not change the communication bounds of $\Omega{(mn)}$ of the current parallel algorithmic approaches.
\begin{proof}
Fragment-Ion index is based on indexing the peaks for each of the theoretical spectra. If the indexing is taking place on the head node then $\Omega{(mn)}$ communication has to take place to distributed the index on each of the processing nodes. 
\end{proof}

\end{corollary}

\begin{theorem}\label{any}
We prove that much tighter lower-bounds are possible for parallel algorithms (that are yet to be discovered). Combining the lower-bounds on $W$, $L$, and $F$ will yield lower-bounds on the overall run time \textbf{possible} for processor with 
$M \leq (mn+\frac{2qr}{p})$ memory available. Therefore, the lower-bounds \emph{possible} for parallel algorithms is equal to $\Omega({\frac{nmq}{p}})$. 
\begin{proof}
Combining the lower-bounds on $W$, $L$, and $F$ will yield lower-bounds on the overall run time of the existing HPC algorithms. In our theorems we have proved that $L =(\frac{2}{mpn^2})$ + $F = \frac{qm(2n-1)}{p}$ +
$W = \frac{m^2}{pm+2q}$. This summation gives us a result of $\Omega({\frac{2mnq}{p}})$.
\end{proof}
\end{theorem}

As can be seen from Theorem \ref{existing} that the existing HPC algorithm achieve only $\Omega{(mn)}$ run time irrespective of the number of processors that are used for the computations. Any advantage that is observed in the experiments are likely due to the smaller subset of spectra $q$ that needs to be processor on each processor. However, with high throughput mass spectrometers $q$ is approaching the theoretical databases, and any advantage is by a constant factor than asymptotic. 

On the other hand, we can see the Theorem \ref{any} predicts $\Omega({\frac{nm^2}{p}})$ as the overall run time possible for database and spectra search when $m$ is approx. equal to $q$. Although estimate of the lower-bound can be done by approximating $q$ to $m$ which allow for much simpler mathematical expressions but overestimates the lower-bound of the run time. In reality the run time is closer to $\Omega({\frac{nmq}{p}})$ which incurs a parameter for the number of spectra as well in the expression. However, with the latest usage of database search algorithms that require more number of post-translations modification parameters, and larger window size; the dominating factor will remain the communication costs related to the theoretical database. 

We specifically note here, that none of the HPC techniques proposed till date achieve this lower-bound of computation and communication. Significant research efforts is needed to ensure that parallel algorithms can be designed which achieve these lower-bounds both in theory and in practice. 

\section{Meta-Analysis of Results of Current HPC methods} \label{meta-analysis}
To confirm our lower-bounds that we have proved for the existing methods, and lower-bounds on communication that might be possible we did a thorough evaluation of the existing methods. These existing methods \cite{10.1093/bioinformatics/btw721,10.1093/bioinformatics/btr523, li2019swpepnovo, diament2011faster, sun2014improved,kulkarni2009scalable,baumgardner2011fast,li2019sw,li2019mctandem,pratt2012mr, bjornson2008x,duncan2005parallel} included MPI-based memory-distributed implementations, Map-Reduce/Hadoop implementations, and GPU-based methods. Since we are assuming a memory-distributed architecture for our bounds; we have concentrated on those studies. Further, we have eliminated studies that have been conducted on a cloud-based Hadoop systems since communication patterns, and infrastructure information is generally not available for commercial or shared facilities. We have also discarded numbers for CPU-GPU based algorithms since it is a distinctly different architecture than a homogeneous memory-distributed machines assumed for our calculations. 

\begin{figure*}[h] 
\centering
\vspace{-2.5cm}
\includegraphics[scale=0.65]{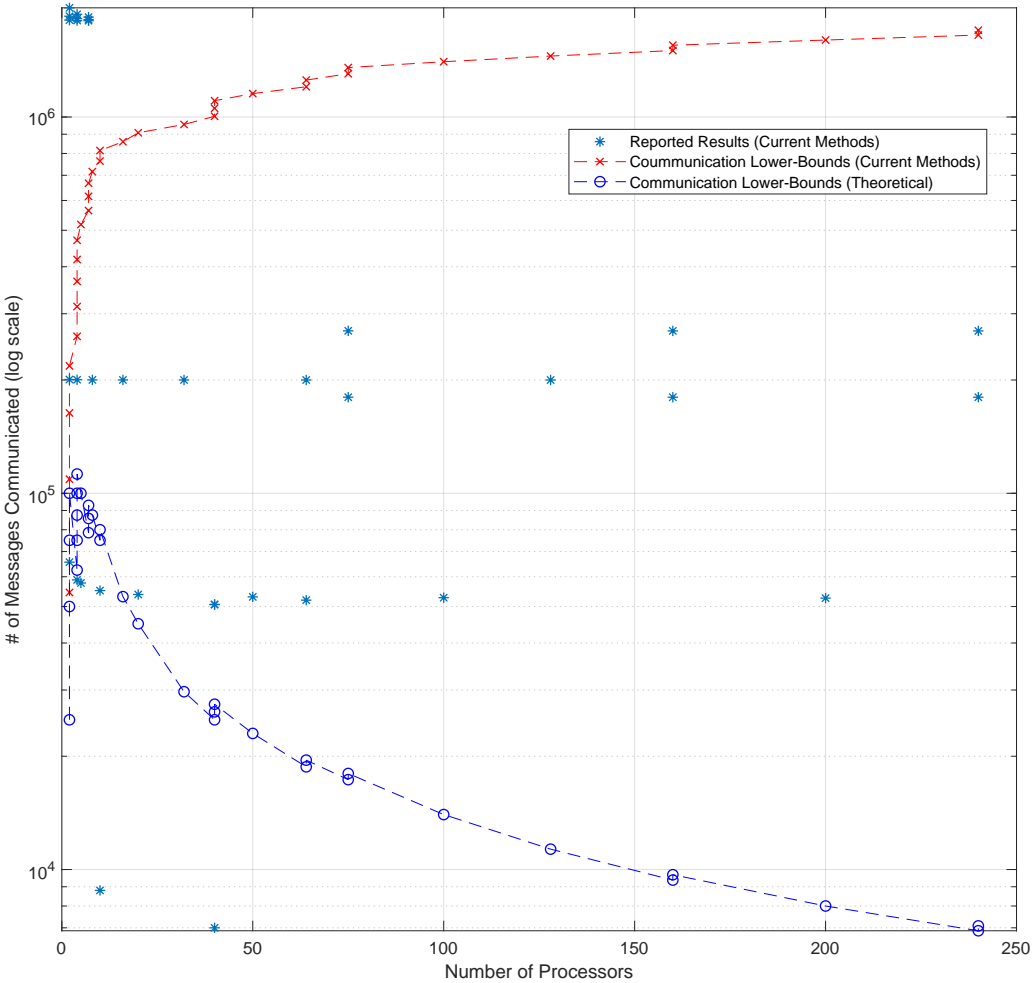}
\vspace{-3.0cm}
\caption{
The graph show the amount of communication that takes place with increasing number of processors. As can be seen that the most of the HPC methods that are listed do not achieve the lower-bounds on the communication. The gap increases rapidly between the communication required for the state-of-the-art HPC algorithms, and the communication that can be theoretically achieved.}
\label{comm-fig}
\end{figure*}

We concentrated on two metrics to make sure that the comparisons are fair for methods that may have been tested on different set of architectures, and systems. One of these metric is the \emph{amount of total communication} for a given parallel algorithm, and this metric is going to be independent of machines, and systems. The second key metric used for estimating the efficiency of these parallel algorithms is \emph{speedups}. Similar to the communication metric, speedups are also independent metric that is not based on comparison with other architectures.

For evaluation, we downloaded all the results \cite{10.1093/bioinformatics/btw721,10.1093/bioinformatics/btr523, li2019swpepnovo, diament2011faster, sun2014improved,kulkarni2009scalable,baumgardner2011fast,li2019sw,li2019mctandem,pratt2012mr, bjornson2008x,duncan2005parallel} that have been reported till date. This information included, the database size, the number of spectra, serial and parallel times, and the speedups. Memory (GB) was also noted whenever reported. Using this information, we plotted the communication message that was required for the method to complete. Note that we only consider the amount of data that needs to be communicated as a function of theoretical database, and neglecting the length of the theoretical spectra. We then plotted the communication bounds that we have calculated for the current methods, as well as the communication bounds that are theoretically possible. As can be seen in Fig. \ref{comm-fig}, that most of the results that are reported are close to the bounds that we have calculated. Also note that as the number of processor increase, the number of messages that need to be transmitted (theoretically) rapidly decrease; however, such behaviour is not exhibited by real-world implementations. Clearly, this is because majority of existing HPC methods do not consider the communication cost in their design. 

We are only aware of this study \cite{kulkarni2009scalable} which allowed splitting the database among parallel nodes. However, as our later analysis shows that the speedups attained by this method is still less than linear. This is because the communication-costs are masked by on-the-fly computations leading to high compute times and limited (around 50\%) parallel efficiency. The study also assumes that the number of spectra are much less than the theoretical database which is no longer valid due to high-throughput mass spectrometers. 

\begin{figure*}[h] 
\centering
\vspace{-6cm}
\includegraphics[scale=0.75]{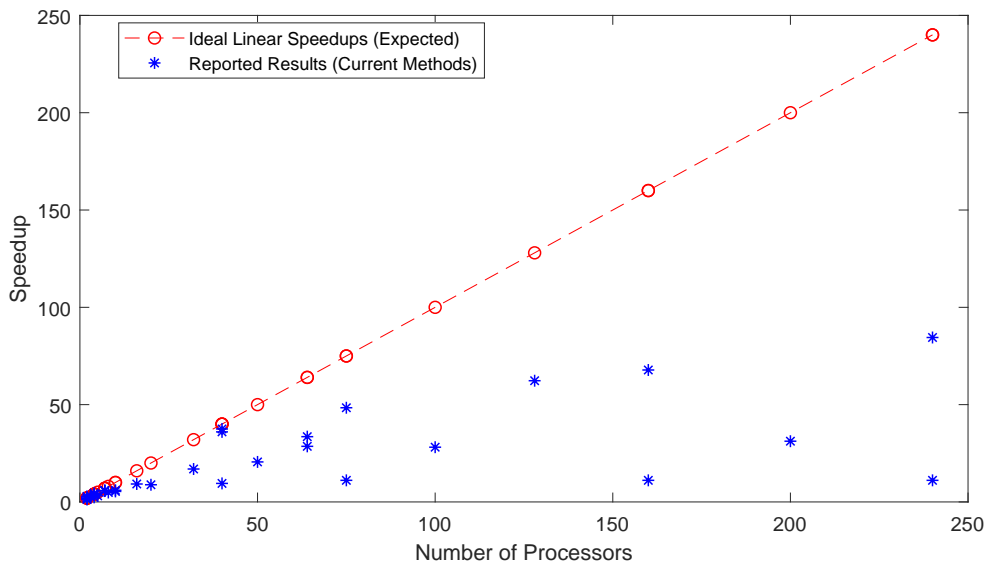}
\vspace{-7cm}
\caption{This graph represents that speedups that are reported by the papers, and the corresponding linear-speedup that can be achieved with increasing number of processors. Note that reported results that are listed here are also the results that are depicted in Fig \ref{comm-fig} and shows a one-to-one correspondence between the amount of communication and the speedups with increasing number of processors}
\label{speed-fig}
\end{figure*}

To validate that our estimates were correct; we went one step further and looked closely into the speedups that were being reported. The speedups that are reported as shown in Fig. \ref{speed-fig} conclusively show that increasing the number of processors decreased the speedups that were obtained for these state-of-the-art methods. The decrease in speeds-up, of course, is due to increase in the communication, and the gap between the current methods, and the theoretical bounds that can be achieved; but are currently not attainable. Thus, the rigor of the prior research suggests that there is significant effort that is needed to investigate parallel algorithms that can achieve the lower-bounds that we have proved, and thus give reasonable performance with increasing number of processors, and data.  

\section{Experimental Evaluation of Current HPC methods} \label{experiments}
We evaluated several existing database peptide search tools including MSFragger \cite{kong2017msfragger}, Comet-MS \cite{eng2013comet}, MSGF+ (MS-PyCloud) \cite{kim2014ms}, \cite{chen2018ms}, X!Tandem (X!!Tandem) \cite{bjornson2007x} and SW-Tandem \cite{li2019sw} in parallel configuration by searching increasing size experimental data against various size custom database search-spaces in both open- (precursor mass tolerance $>$ 100Da) and restricted- (precursor mass tolerance $\leq$ 1Da) search modes. The custom database search-spaces were created by increasingly adding variable post-translational modifications (PTMs) to the Uniprot homo sapiens (UP000005640) database. The experimental datasets were created by splitting the dataset: PXD015890 into 3 subsets each containing: 25\%, 50\% and 100\% of the experimental spectra data. In the first experiment, the 25\% subset was searched against the human database incorporating methionine oxidation (M + 15.99Da) modification in restricted-mode. In the second, third and fourth experiments, the 25\%, 50\% and 100\% experimental subsets were searched against a human database search-space incorporating methionine oxidation (M + 15.99Da), STY-phosphorylation (STY + 79.97Da) and N-Term acetylation (n-term + 42.02Da) in restricted search mode respectively. In the fifth and sixth experiments, the 50\% and 100\% subsets were searched against the custom search space incorporating methionine oxidation (M + 15.99Da) and STY-phosphorylation (STY + 79.97Da) in open-search mode respectively. The three-custom search-spaces  grew to 7.6 million, 76 million and 108 million peptides and variants respectively. The fragment mass tolerance was set to 0.01Da where applicable. The experimental spectra charge range was set between 1 and 4, minimum and maximum precursor mass range between 500 and 5000 amu, and the minimum and maximum peptide length was set to 6 and 46 respectively. The experiments were performed on a cluster machine where each node was equipped with a 16 core processor and 32GB RAM, interconnected with 100 GB/s HDR InfiniBand, also connected to a Lustre-based shared storage system via the same interconnect. 

\begin{figure}[htbp]
\centerline{\includegraphics[width=0.99\linewidth]{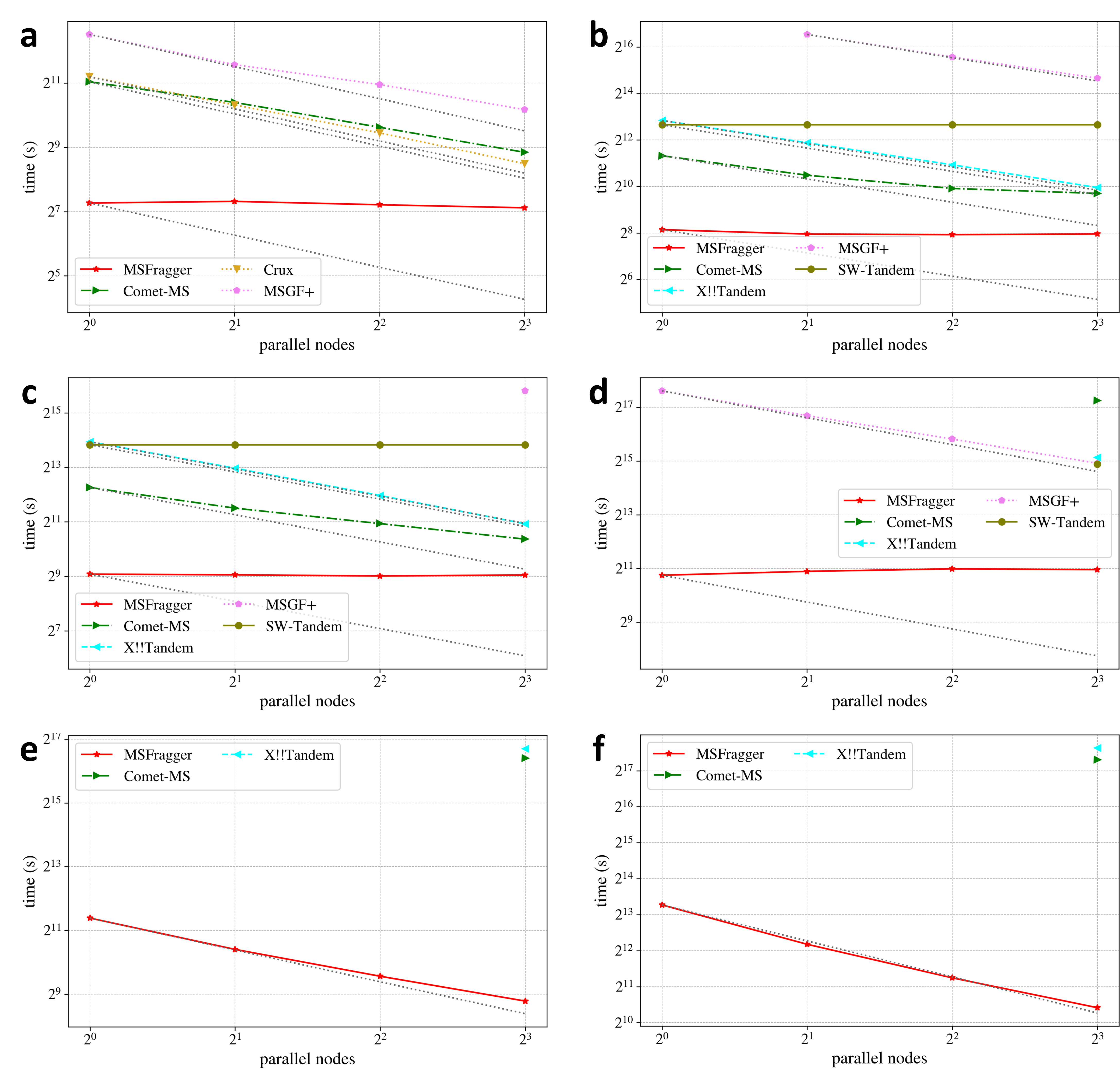}}
\caption{Experimental run times for several tools in log-log scale. The gray dotted lines depict the ideal scalability for each experiment. \textbf{(a to d)} Modern index-based database search tools (such as MSFragger) show a poor parallel performance due to low compute/communication ratio in restricted-search experiment mode. However, it  still performs  faster than older index-free tools such as MSGF+ or X!Tandem. \textbf{(e, f)} The scalability is significantly improved in open-search mode however drops to sub-linear beyond a certain number of parallel nodes due to memory bandwidth saturation.}
\label{fig:scalability}
\end{figure}

\begin{figure}[htbp]
\centerline{\includegraphics[width=0.60\linewidth]{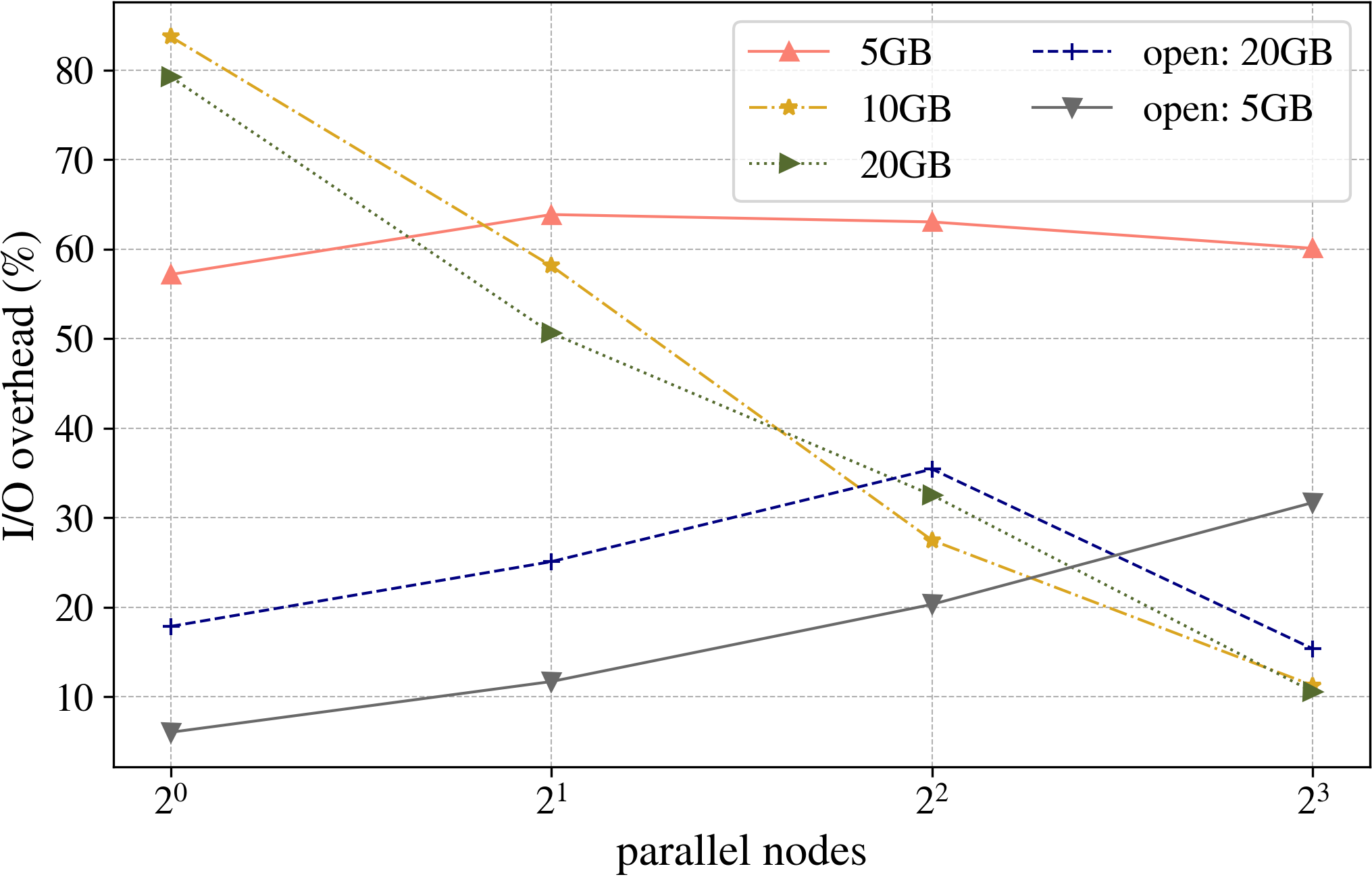}}
\caption{The I/O to overall run time ratio is more than 60\% for MSFragger in restricted-search experiments and drops dramatically as the number of nodes is increased (I/O per node is decreased). In case of open-search, the I/O to compute ratio is much lower.}
\label{fig:io_comm}
\end{figure}

The scalability results depicted in Fig.~\ref{fig:scalability} show that in case of restricted search mode (Fig.~\ref{fig:scalability} a to d), the search tools depict lower scalability than the linear (the positive deviation from the dotted gray line depicting ideal scalability) as most of the time is spent in I/O and data communication with minimal time spent in performing the computations. In open-search mode, MSFragger depicts near linear scalability until a certain number of parallel nodes but drops to sub-linear beyond that point. The reason for this is the poor parallelization technique employed within existing HPC tools (replicate the entire database on all nodes and partition the experimental data among them) which results in higher communication overheads due to memory bandwidth exhaustion. Fig.~\ref{fig:io_comm} further depicts the percentage total time for MSFragger spent in I/O showing that in case of restricted-search mode, the I/O time dominates the parallel performance whereas in open-search mode the I/O time percentage drops. Note that the load imbalance increases dramatically in case of open-search for MSFragger, which can further negatively impact the overall performance. Finally, we confirmed by profiling MSFragger using Intel VTune in open-search mode that its performance is heavily ($>$70\%) memory-bandwidth bounded as the custom database search-space size increases. Tools that do not take advantage of modern  indexing strategies such as Crux, Comet-MS, MSFG+ and X!Tandem perform orders-of-magnitude slower than MSFragger.

\section{Discussions} \label{discussions}
There is an urgent need for \emph{provably} scalable parallel algorithms for large-scale MS systems biology studies which has direct impact on personalized nutrition, microbiome research, and cancer therapeutics. This is especially true for non-model proteomics, meta-proteomics, and proteogenomic studies where the search-space traversal needed to make peptide deductions are massive. Our theoretical results indicate that further formal design, and evaluation is warranted for scalable infrastructure for MS based omics database-workflows. In order to make progress, the next generation of parallel algorithms will have to acquire \emph{provably} demonstrated superior performance on multicore, GPU, memory-distributed supercomputers, and cloud-computing infrastructure. Such contributions are expected to be significant because it will open up novel, and faster ways to analyze MS data for various omics (read: preteomics, proteogenomic, meta-proteomics etc.) studies considered “too large-scale”. Following are few points that would help the reader interpret the theoretical results in this paper: 

\begin{enumerate}
\item For the purposes of this paper we have assumed a single parallel computing strategy for deducing peptides. We do realize that the HPC methods that have been proposed till date have variation such as scoring, getting the candidate theoretical spectra etc. However, the parallel strategy that is used by these HPC methods is similar (as described in the \ref{general-hpc} section) and we are estimating the communication lower-bounds of these parallel algorithms. Since the data is managed in the same way for all of the HPC methods; variations (including theoretical spectra generation) will only modify some constants in these communication bounds. 
\item We further show that the pre-dominant way of proteomics algorithms to increase efficiency by reducing the number of computations (using mass filtering or filtering using other characteristics of mass spectrometry data) does not change the communication-bounds that are being depicted by \emph{current} state-of-the-art parallel algorithms. However, we also show that parallel algorithms with much tighter bounds are possible (but are not yet discovered). 
\item We design and implement parallel computing solutions for problems that are compute- or memory-intensive. Further, such parallelization is accomplished when the problem is not scalable for a single node i.e. it is very large in data or computations. Note that communication-bounds that we have proved are with the assumption that the theoretical database (or spectra that needs processing) are very large and do not fit in memory M of single machine. If the size of the data is not that large (i.e. all database and spectra are fitting in a Memory M) then parallelising will result in speedups that may be expected to be larger than the bounds that we just proved. However, these results and speedup will just be a artifact of the system and/or data being analyzed and will not be a generalizable result. That is why we repeatedly see that adding more number of processor do not significantly scale the computations and the experimental results that are published are for relatively small datasets. 
\item For the current bounds we have assumed that the theoretical database is on the master node and is communicated via the network. However, if the whole database is not communicated (e.g. only if database sequences are communicated), then the amount of communication is substituted by computation costs that would be needed for further computations i.e. $O(nm^2/p)$. Therefore, the lower-bounds that are achieved by the current HPC methods still hold true. This is also confirmed by the meta-analysis of HPC methods published results.
\item For calculating our bounds we assume that whole database is needed for computations. One can argue that 'candidate spectra' are the only real-computations that are done by the algorithms. This reasoning also does not effect the lower-bounds that are calculated. The reason is that having 'candidate-spectra' \emph{does} reduce the amount of computations. However, we have shown that the amount of communication is the real bottleneck for these parallel algorithms. Since calculation of candidate-spectra still requires access, and communication of the theoretical spectra-database; the communication bounds (i.e bottleneck) remains unchanged even when only candidate-spectra are used for computations.
\end{enumerate}

\section{Conclusions} \label{conclusions}
For the past 30 years, significant efforts have been invested for the design, and development of efficient search engines for MS based omics data analysis. These methods are numerical algorithms developed for MS based peptide deduction, and are designed by assuming arithmetic operations as the sole metric for efficiency. In the last decade, the technological trend of the Moore's law has kept making the arithmetic operations faster. As a result, bottleneck for many MS algorithms have shifted from computational arithmetic operations efficiency to communication of data between different levels of memory-hierarchy or between different nodes in a distributed-memory architecture. This bottleneck has resulted in unusually long processing times even for high-performance computing algorithms. However, the poor scalability of these MS based omics algorithms has been considered an artifact of the data, or the architectures, and have been subjective and anecdotal, till date. 

In this paper, we formulate, and quantify the efficiency of the current state of the art HPC algorithms for MS data analysis. We have presented and proved lower bounds on the amount of \emph{communication} that is achieved by the current MS based omics HPC methods. We also prove the lower-bounds that \emph{can} be achieved by parallel algorithms on a distributed-memory architecture. To the best of our knowledge, this is the first study to formulate a theoretical framework showing that the existing parallel strategies for MS based omics data analysis are not achieving the communication bounds that are possible, and that continued improvements are needed in this area of research. Meta-analysis of existing literature agrees with our theoretical analysis i.e. sup-optimal communication costs are achieved by existing MS based omics HPC tools. The experiments that we have performed also concur with these bounds. Therefore, novel parallel algorithms that exhibit optimal-communication costs are needed that can close the communication gap between theory, and practice for MS based omics algorithms. Improved design, development, and implementation of such communication-avoiding parallel algorithms will allow computations of MS based proteomics, meta-proteomics, and proteogenomic data that could scale gracefully with increasing number of processors. 

In contrast to existing methods, the next generation of HPC algorithms must be designed by considering \emph{both} computational, and communication costs as metrics for efficiency. These designs will allow us to experiment by varying balance points between communication, and computation, and with different computing platforms including distributed-memory clusters, supercomputers, and commodity AWS cluster which is primarily used for cloud-computing. We assert that next generation of parallel algorithms that can scale (at least) linearly with increasing number of processors, size of the (theoretical) database, and spectra will be essential for scalable MS omics studies. This will be especially important for machine-learning, and deep-learning MS methods that are notorious for scalability bottlenecks.

\section{Acknowledgements}
The author would like to thank Usman Tariq, and Fatima Afzali for their useful comments, and suggestions. Research reported in this paper was supported by NIGMS of the National Institutes of Health under award number: R01GM134384. The content is solely the responsibility of the authors and does not necessarily represent the official views of the National Institutes of Health. Fahad Saeed was further supported by the National Science Foundations (NSF) under the Award Numbers NSF CAREER OAC-1925960. The content is solely the responsibility of the authors and does not necessarily represent the official views of the National Science Foundation.

\newpage
\bibliographystyle{unsrt}
\bibliography{references}

\end{document}